# Real-space study of zero-field correlation in tetralayer rhombohedral graphene

Yufeng Liu[1,2,†], Zonglin Li[1,2,†], Shudan Jiang[1,2,†], Min Li[3], Yu Gu[1,2], Kai Liu[1,2], Qia Shen[1,2], Liang Liu[1,2,6], Xiaoxue Liu[1,2,6], Dandan Guan[1,2,6], Yaoyi Li[1,2,6], Hao Zheng[1,2,6], Canhua Liu[1,2,6], Kenji Watanabe[4], Takashi Taniguchi[5], Jinfeng Jia[1,2,6], Tingxin Li[1,2,6], Guorui Chen[1,2,6], Jianpeng Liu[3\*], Can Li[1,2\*], Zhiwen Shi[1,2,6\*], and Shiyong Wang[1,2,6\*]

[1]Tsung-Dao Lee Institute, Shanghai Jiao Tong University, Shanghai, 201210, China

[2]Key Laboratory of Artificial Structures and Quantum Control (Ministry of Education), School of Physics and Astronomy, Shanghai Jiao Tong University, Shanghai 200240, China

[3]School of Physical Science and Technology, ShanghaiTech Laboratory for Topological Physics, ShanghaiTech University, Shanghai 201210, China

[4]Research Center for Electronic and Optical Materials, National Institute for Materials Science, 1-1 Namiki, Tsukuba 305-0044, Japan

[5]Research Center for Materials Nanoarchitectonics, National Institute for Materials Science, 1-1 Namiki, Tsukuba 305-0044, Japan

[6]Hefei National Laboratory, Hefei 230088, China

[†]These authors contribute equally to this work.

[\*]Emails: liujp@shanghaitech.edu.cn, lic_18@sjtu.edu.cn, zwshi@sjtu.edu.cn, shiyong.wang@sjtu.edu.cn

**Rhombohedral graphene (RG) has emerged as a promising platform for exploring exotic quantum phenomena, such as quantum magnetism, unconventional superconductivity, and fractional quantum anomalous Hall effects. Despite its potential, atomic-scale investigations of RG remain limited, hindering a detailed microscopic understanding of the origins of these correlated states. In this study, we employ scanning probe microscopy and spectroscopy to probe the intrinsic electronic states in trilayer and tetralayer RG. We identify a correlated insulating state with a 17 meV gap at the charge neutrality point in tetralayer RG, which is absent in the trilayer configuration. This gap is suppressed by applying a perpendicular magnetic field or doping the charge carrier density and does not exhibit inter-valley coherence patterns. We attribute this phenomenon to a symmetry-broken layer antiferromagnetic state, characterized by ferrimagnetic ordering in the outermost layers and antiferromagnetic coupling between them. To further investigate this magnetic correlated state, we conduct local scattering experiments. Within the correlated regime, a bound state emerges around a non-**

**magnetic impurity but is absent near magnetic impurities, suggesting that non-magnetic doping induces a spin texture in the ferrimagnetic surface layers. Outside the correlated regime, Friedel oscillations are observed, allowing precise determination of the band dispersion in tetralayer RG. These findings provide atomic-scale evidences of zero-field correlations in RG and may be extended to study other exotic phases in RG.**

Rhombohedral graphene exhibits a unique stacking order in which each successive layer is shifted by one atomic lattice relative to the layer below, resulting in a crystal structure with broken inversion symmetry and a pair of nearly flat electronic bands near the charge neutrality point[1–4]. The energy dispersion of these flat bands evolves with the number of layers, following a power-law dispersion relation $E \propto P^N$, where $E$ is energy, $P$ is momentum, and $N$ is the number of atomic layers[4]. This evolution transitions from a parabolic band structure in bilayers to increasingly flatter bands in thicker multilayers, enhancing electron-electron interactions as $N$ increases and driving the emergence of symmetry-breaking correlated states [5–31].

A correlated insulating state at the charge neutrality point (CNP) and zero magnetic field, often referred to as zero-field correlation, has been observed in tetralayer, pentalayer, and multilayer RG on hexagonal boron nitride (hBN) substrates[13,26,27,32,33]. However, this state is absent in bilayer and trilayer RG systems under similar conditions, as demonstrated by transport measurements. Interestingly, in suspended rhombohedral bilayer and trilayer graphene, zero-field correlations have been observed, attributed to reduced substrate screening and enhanced electron-electron interactions[6,7,24,31,34]. These findings highlight the critical role of environmental factors in governing correlated electronic phenomena. A variety of symmetry-broken correlated states have been proposed, including the quantum valley Hall state, the quantum anomalous Hall state, the quantum spin Hall state, the layer antiferromagnetic state, and the inter-valley coherence excitonic insulating state[6,23,25,26,30,31,34–36]. However, these phenomena are highly sensitive to experimental conditions such as substrate-induced screening, atomic lattice reconstruction, interlayer pressure, and perturbations from measurement probes. This sensitivity complicates the interpretation of experimental data, as results can vary significantly both qualitatively and quantitatively across different techniques and setups. To date, the lack of consensus on the exact nature of the zero-field correlated ground states underscores the importance of real-space studies. Such techniques are critical for providing direct, atomic-scale insights into the spatial distribution of electronic correlations, lattice reconstructions, and symmetry-breaking orders[37–39].

In this study, we performed gate-tunable scanning tunneling microscopy and spectroscopy (STM/STS) investigations of RG trilayers and tetralayers on hBN at 1.3 K/4.2 K. A pair of flat bands was observed in both systems, with the bandwidth and band separation tunable via an applied perpendicular electric field. The local density of states (LDOS) in the tetralayer is significantly stronger than in the trilayer, facilitating the emergence of a zero-field correlated insulating state with a 17 meV gap in the tetralayer. This correlated gap is suppressed by applying a perpendicular magnetic field or by doping the charge carrier density. We attribute this correlated insulating state to a symmetry-broken layer antiferromagnetic (LAF) phase, characterized by ferrimagnetic ordering in the outermost layers and antiferromagnetic coupling between them. Importantly, we rule out the possibility of an intervalley coherence insulating state, as no inter-valley scattering patterns were detected. To further investigate the magnetic properties of this state, we performed local scattering experiments, which revealed that the outermost surface exhibits characteristics of a two-dimensional quantum magnet. Within the correlated regime, a bound state forms around a non-magnetic impurity but is absent near magnetic impurities, indicating that non-magnetic doping induces a spin texture in the ferrimagnetic surface layers. These findings provide direct real-space evidence of the symmetry-broken nature of the zero-field correlated state and its magnetic properties.

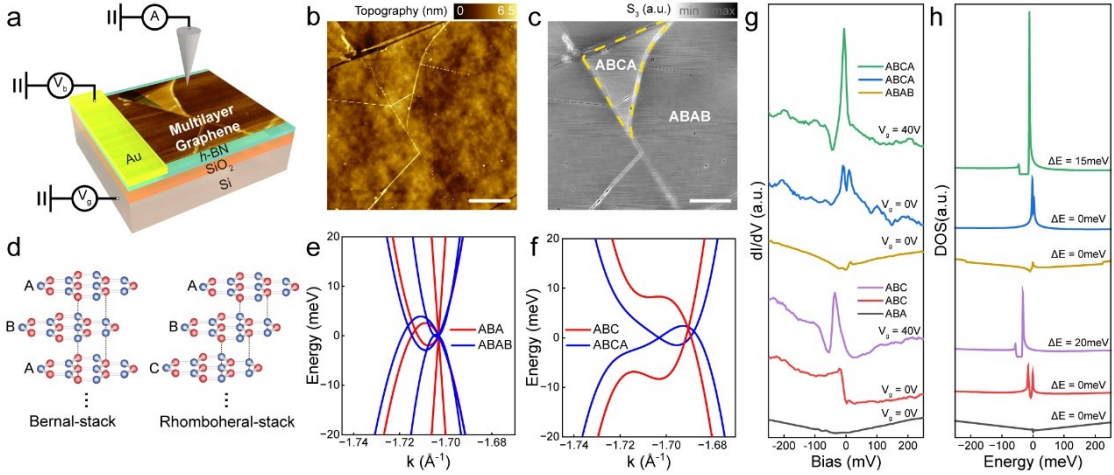

**Fig. 1. Device Setup and Flat Bands in Trilayer and Tetralayer Rhombohedral Graphene. a,** Schematic of a gate-tunable device for STM/STS measurements. **b,c,** AFM and SNOM images of tetralayer graphene. The bright areas indicate metastable rhombohedral (ABCA) stacking stabilized by regions of stable Bernal (ABAB) stacking. Scale bar: 1um. **d,** Schematic representation of rhombohedral and Bernal stacking arrangements. **e,f,** Calculated band structures near the K points for ABA, ABAB, ABC, and ABCA graphene multilayers. **g,** Differential conductance (dI/dV) spectra of graphene multilayers with different stacking orders, measured with and without gate voltage. **h,** Tight-binding calculations of the density of states for graphene multilayers, with and without an applied electric field.

## Electronic structure of few-layer graphene with various stacking orders

To explore the flat bands in multilayer RG, we fabricated gate-tunable STM devices, as illustrated in Fig. 1a. The multilayers were assembled using the standard tear-and-stack method, allowing precise control over the stacking order. Rhombohedral stacking was verified using scanning near-field optical microscopy (SNOM), which clearly distinguished between different stacking configurations (details of the fabrication process are provided in Methods and Extended Fig. 1-3). Atomic force microscopy (AFM) imaging, shown in Fig. 1b, reveals uniform height across the sample, while SNOM imaging highlights optical contrast variations corresponding to different stacking orders. Noted that the metastable ABCA can easily transition into the energetically favorable ABAB stacking. Our AFM and SNOM imaging further demonstrate that the metastable ABCA stacking can be stabilized by domain boundaries, indicated by dashed white lines in the image. These boundaries serve as structural constraints, preserving the rhombohedral configuration and preventing spontaneous relaxation to Bernal stacking.

Figure 1e and f illustrate the band structures of four distinct stacking configurations: ABA, ABC, ABAB, and ABCA. Near the charge neutrality point (CNP), the bands in ABA and ABAB stackings are dispersive, whereas those in ABC and ABCA stackings are relatively flat, characteristic of rhombohedral configurations. Low-temperature STM and STS measurements, presented in Fig. 1g, provide structural and electronic information of trilayer and tetralayer graphene (Extended Fig 2-3). Near the CNP, the ABC and ABCA stackings exhibit significantly stronger LDOS intensities compared to the ABA and ABAB stackings, confirming the presence of flat bands in the rhombohedral configurations. When a perpendicular electric field is applied via the back gate, a bandgap emerges, and the band edges become distinctly flatter. This is accompanied by an enhanced LDOS at the band edge, consistent with tight-binding calculations (Fig. 1f and extended Fig. 4). Notably, the ABCA stacking at the CNP shows a higher LDOS intensity and a narrower band width than ABC stacking (7 meV vs. 10 meV), revealing the layer-dependent evolution of band dispersion in RG.

**Electric field-dependent electronic structure of trilayer Rhombohedral graphene**

The band structure of multilayer RG is highly sensitive to an applied electric field[3]. To explore this dependence, we first examined the influence of a back-gate electric field on the band structure. A key challenge in these measurements was minimizing the impact of the STM tip, which can introduce a significant local electric field due to the work function difference between the tip and the sample. To address this, we utilized a sharp copper tip (Fig. 2a) with a work function closely matching that of graphene, ensuring accurate measurements of the intrinsic electronic structure. As shown in Fig. 2b, the intrinsic electronic structure evolves systematically with increasing back-gate electric field, exhibiting a gradual opening of the band gap. At positive gate voltages, the LDOS is enhanced at the conduction band (CB) minimum, while at negative voltages, it is enhanced at the valence band (VB) maximum. Notably, no correlated gap was observed at the charge neutrality point, consistent with transport measurements. These experimental results align closely with single-particle tight-binding calculations, demonstrating the role of field-tunable band structures in RG multilayers. Figure 2c illustrates calculated electric-field-dependent and layer-resolved density of states (DOS) spectra. The low-energy flat bands are dominated by LDOS contributions from the outermost layers, which are particularly sensitive to the applied electric field. Since STM primarily probes the topmost layer, the measured spectra closely match the calculated DOS of the surface layer, confirming that STM can accurately capture the layer-specific electronic properties of RG without perturbing its intrinsic electronic structure.

Next, we investigated the influence of the STM tip-induced electric field on the electronic structure of trilayer RG. The use of a gold STM tip introduces a pronounced local electric field to the top graphene layer due to the work function mismatch between gold and graphene. This field can significantly modify the local electronic environment, providing an opportunity to study tip-dependent electronic phenomena in trilayer RG. Figure 2d displays data collected under the influence of this tip-induced electric field, revealing an asymmetric band gap opening that depends on the polarity of the applied back-gate electric field. This asymmetry arises from the tip-induced field and is reproduced by tight-binding calculations including a potential energy (+20meV) on the topmost layer, as shown in Fig. 2e. Beyond the asymmetric gap opening, we observed a divergence in the chemical potential when the Van Hove singularity crossed the Fermi level. This is manifested as a distinct "kink" feature in Fig. 2e,f near zero back-gate voltage. This feature underscores the sensitivity of the electronic structure to the local density of states, particularly near critical points in the band structure.

At the CNP, we detected a large insulating gap of 45 meV (Extended Fig. 5). The size of this insulating gap at the CNP exhibits strong sensitivity to the condition of the STM tip, suggesting that it may not necessarily be attributable to many-body correlation effects. Instead, it appears to arise from extrinsic factors, such as tip-induced fields and local electronic perturbations. This finding underscores the complexity of disentangling intrinsic and extrinsic effects in STM measurements. These results may reconcile discrepancies among previous STM studies on few-layer graphene[14,30,40], where reported gap sizes and the nature of correlated states have varied widely. Such variability likely reflects differences in experimental conditions, particularly the influence of the STM tip as well as inhomogeneous substrate. Our findings highlight the critical importance of mitigating tip-induced effects to accurately probe the intrinsic electronic properties of graphene multilayers, as the electronic structure and correlated states in RG are highly sensitive to the STM tip's work function, the magnitude and direction of local electric fields, and substrate conditions.

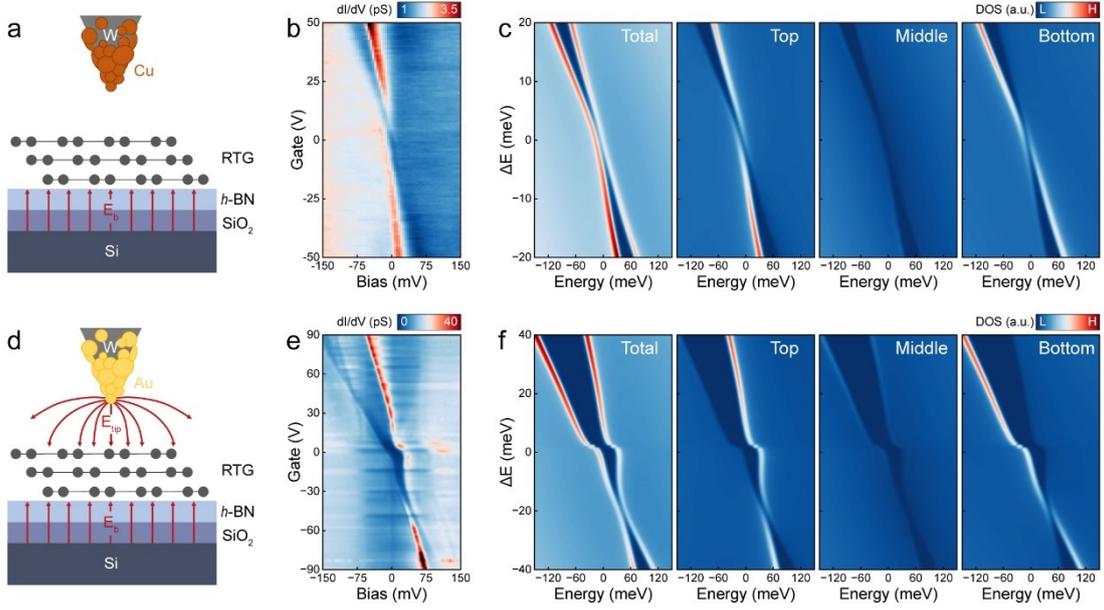

**Fig. 2. Electric field-dependent electronic structure of trilayer RG**. **a,** Schematic of the trilayer RG setup using a copper (Cu) STM tip. A silicon back-gate is employed to apply a perpendicular electric field. **b,** $dI/dV$ spectra as a function of back-gate voltage, illustrating the evolution of electronic structure under varying electric fields. **c,** Tight-binding calculated DOS spectra under different interlayer potential differences. From left to right: total DOS, top-layer DOS, middle-layer DOS, and bottom-layer DOS, respectively. **d,** Schematic of the trilayer RG setup using a gold (Au) STM tip, where the tip introduces an additional top electric field. **e,** $dI/dV$ spectra as a function of back-gate voltage, showing an asymmetric gap opening caused by the tip-induced electric field. **f,** Tight-binding calculated DOS spectra including a potential energy (+20meV) on the topmost layer, illustrating the asymmetric impact of the top electric field. From left to right: total DOS, top-layer DOS, middle-layer DOS, and bottom-layer DOS, respectively.

## Zero-field correlation in tetralayer RG

We conducted a detailed investigation into the intrinsic electronic structure of tetralayer RG, minimizing the influence of the STM tip (Fig. 3). Figure 3a presents the *dI/dV* spectra as a function of back-gate voltage using a copper tip. The flat bands exhibit behavior similar to those in trilayer RG, including the opening of a band gap and increased band edge flatness under an applied back-gate electric field. These trends are well reproduced by single-particle tight-binding calculations, where the calculated top-layer DOS spectra align closely with experimental results (Fig. 3c and Extended Fig. 6). This agreement indicates that the measurements accurately reflect the intrinsic electronic properties of tetralayer RG, with minimal influence from the tip-induced electric field.

Unlike trilayer RG, tetralayer RG displays a distinct gap feature near the CNP, which cannot be captured by tight-binding calculations (highlighted by dashed box in Fig. 3a). To better visualize this gap, we performed high-resolution measurements with a smaller modulation amplitude of 0.8 meV at 1.3K (Fig. 3b). The insulating gap is more clearly visible, with a maximum size of 17 meV at $V_g$=0. This gap diminishes rapidly with charge carrier doping induced by the back-gate electric field. STM imaging conducted both within and outside the gap (Fig. 3d) revealed no inter-valley scattering patterns, ruling out the possibility that the observed gap corresponds to an intervalley coherence excitonic insulator, which would manifest as inter-valley scattering patterns in STM imaging. To further exclude local defects or intravalley trigonal warping-induced charge order states as the origin of the gap, we recorded a 500 nm-long *dI/dV* line map (Fig. 3f). The line map shows that the correlated gap is nearly homogeneous across the surface, supporting the interpretation that it arises from an intrinsic correlated ground state. Notably, these findings were corroborated by measurements from an independent tetralayer RG device, as shown in Extended Fig. 7.

To investigate the nature of this correlated state, we further investigated the dependence of the correlated gap on an applied perpendicular magnetic field (Extended Fig. 8). As shown in Fig. 3g, the insulating gap decreases progressively with increasing magnetic field strength. From these measurements, we extracted an effective Landé *g*-factor up to 35, which decreases with charge carrier doping (back-gate voltage). This unusually large Landé *g*-factor indicates that the insulating ground state is valley-polarized, rather than originating from a spin-imbalanced ground state. A valley-polarized state implies that the correlated insulating phase is primarily governed by valley symmetry breaking, which is sensitive to the interplay between the applied magnetic field and the intrinsic electronic correlations in the system. Such observations suggest that the ground state is a valley layer antiferromagnetic (LAF) phase, as suggested by mean-field calculations[41,42]. This state is characterized by ferrimagnetic order on the outermost layers, coupled antiferromagnetically across the structure. The physics of this system resembles the Su-Schrieffer-Heeger model, where tetralayer RG can be approximated as close-packed 1D polyacetylene octamers (Fig. 3e). In this analogy, spin-up density localizes at one end of the chain, while spin-down density appears at the opposite end; In the topmost layers, valley-imbalance gives rise to a 2D ferrimagnetic order.

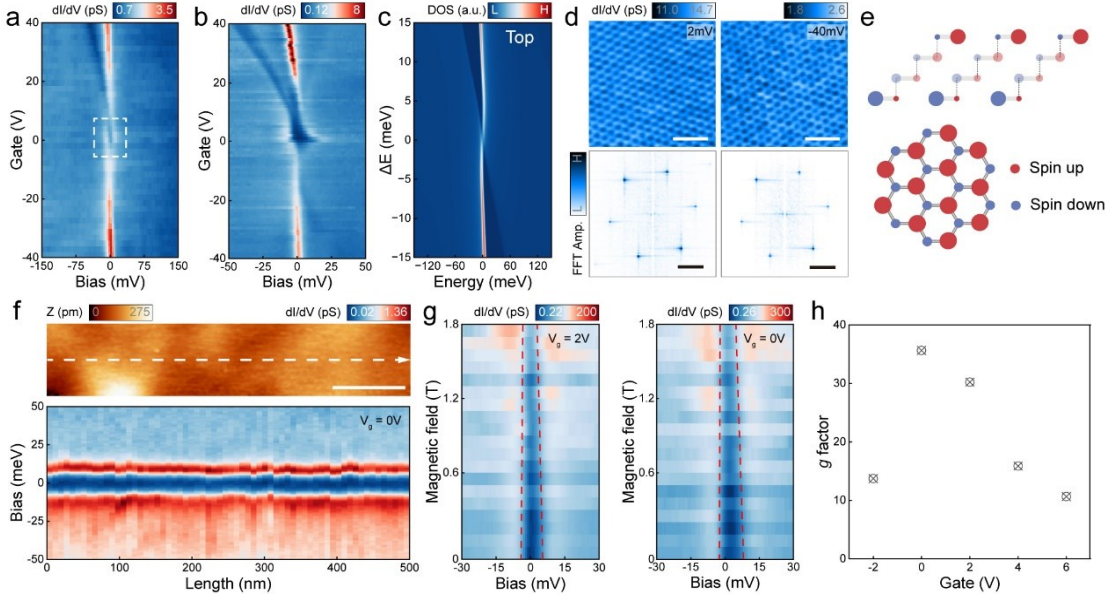

**Fig. 3. Zero-field correlations in tetralayer RG**. **a,** $dI/dV$ spectra as a function of back-gate voltage. **b,** Tight-binding calculated DOS spectra of the top layer. **c,** Zoomed-in $dI/dV$ spectra as a function of back-gate voltage, revealing a correlated gap at the charge neutrality point. **d,** STM images and corresponding FFT image conducted inside and outside the correlated gap, showing no inter-valley scattering patterns. Scale bar:1nm and 2.5nm$^{-1}$. **e,** Schematic depiction of the layer antiferromagnetic ground state, characterized by ferrimagnetic order on the outer layers and antiferromagnetic coupling between them. **f,** Line $dI/dV$ spectra acquired along the dashed line in the STM image above at 0.2T, illustrating a nearly homogeneous correlated gap across the surface. Scale bar: 100nm. **g,** $dI/dV$ spectra as an out-of-plane magnetic field. The dashed line indicates the gap edge. **h,** Extracted Landé $g$-factor as a function of back-gate voltage.

**Friedel Oscillation and localized bound states around defects in RG tetralayer**

The nature of this correlated LAF state is further investigated through real-space scattering experiments. STM imaging enables the direct visualization of the interplay between defects and the electron reservoir by resolving real-space quasiparticle scattering patterns. To systematically study this effect, we deliberately introduced surface defects in tetralayer RG using hydrogen plasma treatment (see Methods for details). Figure 4a presents a typical STM image of the post-treatment RG surface, revealing two distinct types of defects, highlighted within red and blue dashed squares and referred to as type I and type II defects, respectively. We attribute type I defects to magnetic defects, likely arising from carbon vacancies or hydrogen-passivated $sp^3$-hybridized carbon atoms induced by hydrogen plasma. These defects locally break the sublattice symmetry of graphene, resulting in a net magnetic moment. In contrast, type II defects are nearly invisible in STM topographic imaging but exhibit a strong potential to scatter surface quasiparticles. The subtle nature of type II defects suggests they may originate from local imperfections or disorders in the underlying boron nitride substrate. As shown in Fig. 4b, such assignments are further confirmed by zoomed-in STM imaging, revealing inter-valley scattering patterns near type I defects by sublattice symmetry-breaking effects. These patterns are absent near type II defects, further supporting the hypothesis that type II defects act primarily as potential scatterers without inducing magnetic moments.

In our experimental setup, we utilized a gate-tunable approach to control the back-gate electric field, allowing the system to transition between a correlated LAF phase and a normal metallic phase. In the normal metallic phase, standing-wave-like patterns were observed around both type I and type II impurities, indicative of Friedel oscillations[43,44] (Fig. 4d). These oscillatory patterns originate from intravalley scattering, with their wavelengths directly linked to the Fermi wavevector $k_F$. By analyzing the periodicity of these oscillations across varying bias and gate voltages, we were able to extract the energy-momentum relationship, effectively mapping the band dispersion of RG multilayers (Extended Fig. 9). The extracted dispersion near the charge neutrality point aligns with tight-binding calculations (Fig. 4h).

When the system is tuned into the correlated LAF phase, the scattering patterns around type I and type II defects exhibit strikingly different characteristics. Around magnetic type I defects, we observe ring-like features with intensity decaying quickly near the CNP, indicative of a significant reduction in quasiparticle lifetime in the correlated regime (Fig. 4f). This rapid decay underscores the suppression of coherent quasiparticle states due to the strong electronic correlations, characteristic of the correlated LAF phase. In contrast, type II defects, which are possible non-magnetic, display bound states with a striking three-fold symmetric pattern near the CNP (Fig. 4g). This intriguing behavior may arise from the formation of magnetic bound states induced by non-magnetic impurities in the 2D quantum magnetic environment of the LAF phase. The outermost surface of tetralayer RG in this phase hosts a 2D ferrimagnetic state. Magnetic defects (type I) do not typically generate magnetic bound states because their moments prefer to align with the ferromagnetic order within their respective layers, preserving the system's spin polarization without causing significant local magnetic disruption. This alignment minimizes energy and prevents the formation of localized spin textures. Non-magnetic defects (Type II), however, introduce local potential perturbations that disrupt the delicate balance of the spin order. This disruption can create localized magnetic bound states with spin textures, particularly in a system with strong magnetic correlations[45,46]. The three-fold symmetry of the observed bound state pattern could reflect the underlying lattice symmetry and the interplay between the defect potential and the correlated electronic structure. Further investigations using spin-sensitive experimental techniques, combined with theoretical modeling, are essential to unravel the detailed mechanisms driving these distinct scattering behaviors and bound state formations.

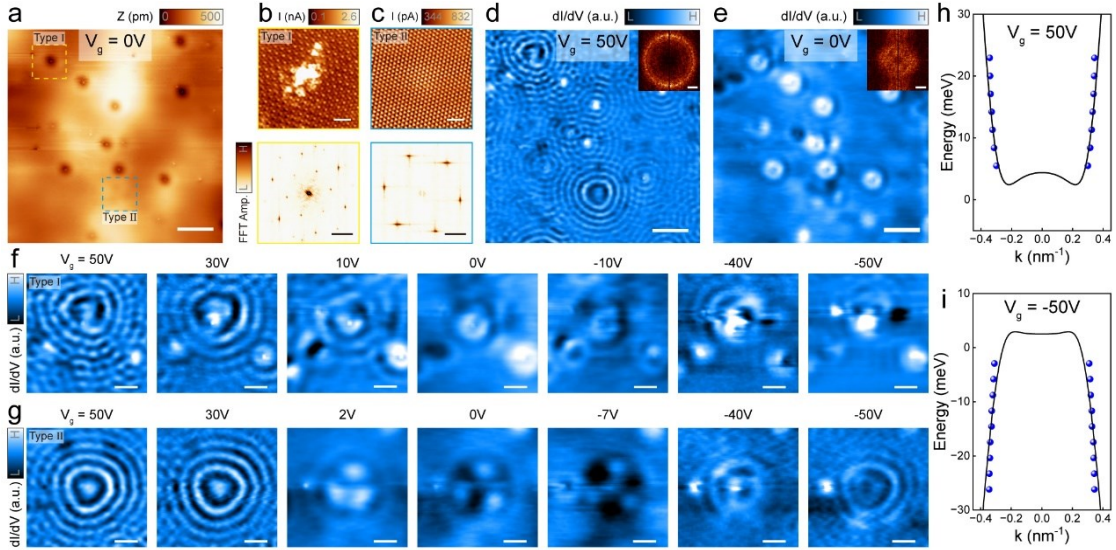

**Fig. 4. Friedel Oscillations and localized bound states around defects in tetralayer RG. a,** Large-scale STM image showing two types of defects. The yellow and blue dashed squares indicate the locations of Type I and Type II defects, respectively. Scale bar: 50nm. **b,c,** Zoomed-in STM images and corresponding fast Fourier transform images around the defects marked in (a). Type I defects break sublattice symmetry, producing inter-valley scattering patterns. Type II defects, in contrast, do not induce such patterns. Scale bar: 1nm and 2.5nm$^{-1}$. **d,e,** dI/dV mappings taken outside and inside the correlated phase. Scale bar: 50nm and 0.045nm$^{-1}$. **f,g,** Zoomed-in dI/dV mappings captured at different back-gate voltages. Scale bar: 20nm. **h,i,** Extracted energy dispersions of the conduction and valence band.

**Conclusions and Outlook**

Through scanning probe microscopy and spectroscopy, we identified a symmetry-broken layer antiferromagnetic ground state in tetralayer RG at the charge neutrality point, featuring a gap up to 17 meV at the CNP. Our findings provide direct atomic-scale evidence of zero-field correlations, revealing the interplay between ferrimagnetic ordering on the surface layers and antiferromagnetic coupling between them. The absence of inter-valley coherence patterns and the distinct scattering behavior near magnetic and non-magnetic defects further elucidate the microscopic mechanisms underlying the correlated phase. These results establish RG as a versatile system for investigating quantum magnetism. Further spin-sensitive and time-resolved experiments could offer deeper insights into the dynamics of the LAF phase and magnetic bound states.

**References**

1. Aoki, M. & Amawashi, H. Dependence of band structures on stacking and field in layered graphene. *Solid State Commun.* **142**, 123–127 (2007).
2. Zhang, F., Sahu, B., Min, H. & MacDonald, A. H. Band structure of ABC-stacked graphene trilayers. *Phys. Rev. B* **82**, 035409 (2010).
3. Koshino, M. Interlayer screening effect in graphene multilayers with ABA and ABC stacking. *Phys. Rev. B* **81**, 125304 (2010).
4. Ho, C.-H., Chang, C.-P. & Lin, M.-F. Evolution and dimensional crossover from the bulk subbands in ABC-stacked graphene to a three-dimensional Dirac cone structure in rhombohedral graphite. *Phys. Rev. B* **93**, 075437 (2016).
5. Winterer, F. *et al.* Ferroelectric and spontaneous quantum Hall states in intrinsic rhombohedral trilayer graphene. *Nat. Phys.* **20**, 422–427 (2024).
6. Velasco, J. *et al.* Transport spectroscopy of symmetry-broken insulating states in bilayer graphene. *Nat. Nanotechnol.* **7**, 156–160 (2012).
7. Bao, W. *et al.* Evidence for a spontaneous gapped state in ultraclean bilayer graphene. *Proc. Natl. Acad. Sci.* **109**, 10802–10805 (2012).
8. Pamuk, B., Baima, J., Mauri, F. & Calandra, M. Magnetic gap opening in rhombohedral-stacked multilayer graphene from first principles. *Phys. Rev. B* **95**, 075422 (2017).
9. Bao, W. *et al.* Stacking-dependent band gap and quantum transport in trilayer graphene. *Nat. Phys.* **7**, 948–952 (2011).
10. Lee, Y. et al. Gate-Tunable Magnetism and Giant Magnetoresistance in Suspended Rhombohedral-Stacked Few-Layer Graphene. Nano Lett. 22, 5094–5099 (2022).
11. Ghazaryan, A., Holder, T., Berg, E. & Serbyn, M. Multilayer graphenes as a platform for interaction-driven physics and topological superconductivity. *Phys. Rev. B* **107**, 104502 (2023).
12. Lu, Z. *et al.* Fractional quantum anomalous Hall effect in multilayer graphene. *Nature* **626**, 759–764 (2024).
13. Zhou, W. *et al.* Layer-polarized ferromagnetism in rhombohedral multilayer graphene. *Nat. Commun.* **15**, 2597 (2024).
14. Zhang, Y. *et al.* Layer-dependent evolution of electronic structures and correlations in rhombohedral multilayer graphene. *Nat. Nanotechnol.* (2024) doi:10.1038/s41565-024-01822-y.
15. Arp, T. *et al.* Intervalley coherence and intrinsic spin–orbit coupling in rhombohedral trilayer graphene. *Nat. Phys.* **20**, 1413–1420 (2024).
16. Guo, Z., Lu, X., Xie, B. & Liu, J. Fractional Chern insulator states in multilayer graphene moiré superlattices. *Phys. Rev. B* **110**, (2024).

17. Dong, J. *et al.* Anomalous Hall Crystals in Rhombohedral Multilayer Graphene. I. Interaction-Driven Chern Bands and Fractional Quantum Hall States at Zero Magnetic Field. *Phys. Rev. Lett.* **133**, (2024).
18. Chatterjee, S., Wang, T., Berg, E. & Zaletel, M. P. Inter-valley coherent order and isospin fluctuation mediated superconductivity in rhombohedral trilayer graphene. *Nat. Commun.* **13**, 6013 (2022).
19. Zhang, H. *et al.* Correlated topological flat bands in rhombohedral graphite. *Proc. Natl. Acad. Sci.* **121**, e2410714121 (2024).
20. Koh, J. M., Alicea, J. & Lantagne-Hurtubise, É. Correlated phases in spin-orbit-coupled rhombohedral trilayer graphene. *Phys. Rev. B* **109**, 035113 (2024).
21. Park, Y., Kim, Y., Chittari, B. L. & Jung, J. Topological flat bands in rhombohedral tetralayer and multilayer graphene on hexagonal boron nitride moiré superlattices. *Phys. Rev. B* **108**, 155406 (2023).
22. Han, T. *et al.* Large quantum anomalous Hall effect in spin-orbit proximitized rhombohedral graphene. *Science* **384**, 647–651 (2024).
23. Kerelsky, A. *et al.* Moiréless correlations in ABCA graphene. *Proc. Natl. Acad. Sci.* **118**, e2017366118 (2021).
24. Myhro, K. *et al.* Large tunable intrinsic gap in rhombohedral-stacked tetralayer graphene at half filling. *2D Mater.* **5**, 045013 (2018).
25. Zhou, H. *et al.* Half- and quarter-metals in rhombohedral trilayer graphene. *Nature* **598**, 429–433 (2021).
26. Liu, K. *et al.* Spontaneous broken-symmetry insulator and metals in tetralayer rhombohedral graphene. *Nat. Nanotechnol.* **19**, 188 (2023).
27. Shi, Y. *et al.* Electronic phase separation in multilayer rhombohedral graphite. *Nature* **584**, 210–214 (2020).
28. Han, T. *et al.* Orbital multiferroicity in pentalayer rhombohedral graphene. *Nature* **623**, 41–47 (2023).
29. Zhou, H., Xie, T., Taniguchi, T., Watanabe, K. & Young, A. F. Superconductivity in rhombohedral trilayer graphene. *Nature* **598**, 434–438 (2021).
30. Hagymási, I. *et al.* Observation of competing, correlated ground states in the flat band of rhombohedral graphite. *Sci. Adv.* **8**, eabo6879 (2022).
31. Weitz, R. T., Allen, M. T., Feldman, B. E., Martin, J. & Yacoby, A. Broken-Symmetry States in Doubly Gated Suspended Bilayer Graphene. *Science* **330**, 812–816 (2010).
32. Han, T. *et al.* Correlated insulator and Chern insulators in pentalayer rhombohedral-stacked graphene. *Nat. Nanotechnol.* **19**, 181–187 (2024).
33. Xie, J. *et al.* Even- and Odd-denominator Fractional Quantum Anomalous Hall Effect in Graphene Moire Superlattices. arXiv.2405.16944 (2024).
34. Seiler, A. M. *et al.* Quantum cascade of correlated phases in trigonally warped bilayer graphene. *Nature* **608**, 298–302 (2022).
35. Lee, Y. *et al.* Competition between spontaneous symmetry breaking and single-particle gaps in trilayer graphene. *Nat. Commun.* **5**, 5656 (2014).

36. Blinov, I. V. *et al.* Partial condensation of mobile excitons in graphene multilayers. arXiv.2303.17350 (2023).
37. Li, G., Luican, A. & Andrei, E. Y. Self-navigation of a scanning tunneling microscope tip toward a micron-sized graphene sample. *Rev. Sci. Instrum.* **82**, 073701 (2011).
38. Jiang, Y. *et al.* Charge order and broken rotational symmetry in magic-angle twisted bilayer graphene. *Nature* **573**, 91–95 (2019).
39. Xie, Y. *et al.* Spectroscopic signatures of many-body correlations in magic-angle twisted bilayer graphene. *Nature* **572**, 101–105 (2019).
40. Liu, Y. *et al.* Visualizing incommensurate inter-valley coherent states in rhombohedral trilayer graphene. arXiv:2411.11163 (2024).
41. Zhang, F., Jung, J., Fiete, G. A., Niu, Q. & MacDonald, A. H. Spontaneous Quantum Hall States in Chirally Stacked Few-Layer Graphene Systems. *Phys. Rev. Lett.* **106**, 156801 (2011).
42. Jung, J., Zhang, F. & MacDonald, A. H. Lattice theory of pseudospin ferromagnetism in bilayer graphene: Competing interaction-induced quantum Hall states. *Phys. Rev. B* **83**, 115408 (2011).
43. Dutreix, C. & Katsnelson, M. I. Friedel oscillations at the surfaces of rhombohedral N-layer graphene. *Phys. Rev. B* **93**, 035413 (2016).
44. Yin, L.-J. *et al.* Imaging Friedel oscillations in rhombohedral trilayer graphene. *Phys. Rev. B* **107**, L041404 (2023).
45. Lin, S.-Z., Hayami, S. & Batista, C. D. Magnetic Vortex Induced by Nonmagnetic Impurity in Frustrated Magnets. *Phys. Rev. Lett.* **116**, 187202 (2016).
46. Park, P. *et al.* Spin texture induced by non-magnetic doping and spin dynamics in 2D triangular lattice antiferromagnet h-Y(Mn,Al)O3. *Nat. Commun.* **12**, 2306 (2021).

**Methods**

**Device fabrications**

All the heterostructures were fabricated via the standard dry transfer method (Extended Data Fig. 1) using a polycarbonate (PC) film placed on a polydimethylsiloxane stamp. Rhombohedral multilayer graphene flakes were obtained by either screening naturally formed rhombohedral domains or twisting graphene layers at a slight angle to facilitate the relaxation of the flakes into rhombohedral stacking. The rhombohedral domains were identified and cut using near-field infrared microscopy and Bruker atomic force microscope respectively. We sequentially picked up layers from top to bottom and released them onto the substrate. Multilayer graphene was pick-up first. Here, if the relaxation method is chosen, a monolayer and a bilayer for ABC stacking, or two bilayer flakes for ABCA stacking, were gradually picked up with a typical small angle of 0.2º. Then a hBN flake was picked up. After stacking, the heterostructures were released onto a Si/SiO$_2$ wafer with pre-deposited electrodes (Ti/Au 10nm/40nm) or onto bare Si/SiO$_2$ substrates, followed by Ti/Au(10nm/40nm) deposition using microfabrication techniques after PC film cleaning. The initial cleaning process removed most of the PC film by sequentially soaking the samples after release with chloroform, acetone, and isopropanol. Then the devices were cleaned with hydrogen plasma at 350°C for 4 hours to ensure the surface was clean. For several devices for which plasma cleaning is insufficient, an atomic force microscope (Park NX7) was used to remove residual contaminants by repeated scanning (contact force ~ 100 nN) in contact mode. Before STM measurement, these devices were annealed in an ultrahigh vacuum chamber overnight at 200°C.

**STM/STS measurements**

STS/STM were carried out using a commercial Unisoku Joule-Thomson STM and a commercial Unisoku 1200JT STM system under low temperature (4.2K/1.3K) and ultra-high vacuum conditions (3 × 10$^{-10}$mbar) and a perpendicular magnetic field up to 3T. The tungsten tips were calibrated using the surface state of a Cu(111)/Au(111) single crystal or through linear I-V curves measured on the Au electrode. A lock-in amplifier (521 Hz, 0.5–5mV modulation) were used to acquire d$I$/d$V$ spectra. The STM tip was navigated to samples using the capacitance-guiding technique[47]. The STM images were processed with WSxM software.

**Theoretical calculation**

We performed tight-binding (TB) calculations to obtain the band structures of ABC-stacked trilayer graphene and ABCA-stacked tetralayer graphene. The full-band TB model incorporates hopping parameters optimized to match local density approximation (LDA) results from density functional theory[48].

The Hamiltonians for these systems are constructed as follows:

For ABC-stacked trilayer graphene:
$$H_{ABC} = \begin{pmatrix} H_{11} & H_{12} & H_{13} \\ H_{21} & H_{22} & H_{23} \\ H_{31} & H_{32} & H_{33} \end{pmatrix},$$

For ABCA-stacked trilayer graphene:

$$H_{ABCA} = \begin{pmatrix} H_{11} & H_{12} & H_{13} & H_{14} \\ H_{21} & H_{22} & H_{23} & H_{24} \\ H_{31} & H_{32} & H_{33} & H_{34} \\ H_{41} & H_{42} & H_{43} & H_{44} \end{pmatrix}.$$

The intralayer Hamiltonian for layers $l = 1, 2, 3, 4$, denoted as $(H_{ll})_{2\times 2}$, is expressed as:

$$H_{ll} = \begin{pmatrix} u_{A_l} + V_{ll} & v_0 \pi^{\dagger} \\ v_0 \pi & u_{B_l} + V_{ll} \end{pmatrix},$$

where $u_{A_l}$ and $u_{B_l}$ are the diagonal site potentials for sublattices $A$ and $B$ in layer $l$. For ABC-stacked trilayer graphene, $u_{A_l} = u_{B_l} = 0$. For ABCA-stacked tetralayer graphene, $u_{A_1} = u_{B_4} = 0$ meV, $u_{B_1} = u_{A_4} = 12.2$ meV, and $u_{A_2} = u_{A_3} = u_{B_2} = u_{B_3} = -16.4$ meV. The term $V_{ll}$ introduces interlayer potential difference ($\Delta E$) caused by an applied perpendicular electric field. These are parameterized as $V = \Delta E(1, 0, 1)$ for ABC-stacked trilayer graphene and $V = \Delta E(\frac{3}{2}, \frac{1}{2}, -\frac{1}{2}, -\frac{3}{2})$ for ABCA-stacked tetralayer graphene.

The interlayer hopping terms, $(H_{ij})_{2\times 2}$ for $i \neq j$, are defined as:

$$H_{12} = \begin{pmatrix} -v_4 \pi^{\dagger} & -v_3 \pi \\ t_1 & -v_4 \pi \end{pmatrix}, \quad H_{13} = \begin{pmatrix} 0 & t_2 \\ 0 & 0 \end{pmatrix},$$

where $t_1$ and $t_2$ represent the interlayer hopping amplitudes between adjacent and next-nearest layers, respectively.

The momentum operator $\pi = vp_x + ip_y$ is defined in terms of the valley index $v = \pm 1$, with $\mathbf{p} = (p_x, p_y)$ measured from the Dirac points $\mathbf{K}_v = \left(v\frac{4\pi}{3a_0}, 0\right)$. Here, $a_0 = 2.46$ Å is the graphene lattice constant.

The Fermi velocity parameters $v_i = \frac{\sqrt{3}a_0}{2\hbar}|t_i|$ for $i = 0, 3, 4$ are derived using the hopping amplitudes $t_i$, where $t_{i=0,1,2,3,4} = (-3100, 356.1, -8.3, 293, 144)$ meV.

To calculate the density of states (DOS) for ABC-stacked trilayer graphene and ABCA-stacked tetralayer graphene, we use the formula:

$$D(\varepsilon) = \frac{1}{N_k} \sum_{n,\mathbf{k}} \delta(\varepsilon - \varepsilon_{n,\mathbf{k}}),$$

where $D(\varepsilon)$ is the DOS at energy $\varepsilon$, $\varepsilon_{n,\mathbf{k}}$ represents the energy of the $n$-th band at wavevector $\mathbf{k}$, and $N_k$ is the total number of $\mathbf{k}$-points evaluated.

The band energies $\varepsilon_{n,k}$ are computed by diagonalizing the tight-binding Hamiltonian over a dense ***k***-point grid in the Brillouin zone. The resulting DOS captures key features of the electronic structure, including the effects of stacking order and interlayer coupling.

## References

47. Li, Guohong et al. "Self-navigation of a scanning tunneling microscope tip toward a micron-sized graphene sample." The Review of scientific instruments 82,7 (2011): 073701.
48. Park, Y. et al. Topological flat bands in rhombohedral tetralayer and multilayer graphene on hexagonal boron nitride moiré superlattices. Phy. Rev. B 108, 155406 (2023).

## Acknowledgement

This work is supported by the National Key R&D Program of China (Nos. 2020YFA0309000, 2021YFA1400100, 2022YFA1405400, 2022YFA1402401, 2022YFA1402404, 2022YFA1402702), the National Natural Science Foundation of China (Nos. 12350403, 12174249, 12174250, 12141404, 92265102, 12374045, 92365302, 22325203, 92265105, 92065201, 12074247, 12174252, 22272050, 21925201, 12304230), the Innovation Program for Quantum Science and Technology (Nos. 2021ZD0302600 and 2021ZD0302500), the Natural Science Foundation of Shanghai (No. 22ZR1430900). T.L. and X.L. acknowledge the Shanghai Jiao Tong University 2030 Initiative Program. X.L. acknowledges the Pujiang Talent Program 22PJ1406700. T.L. acknowledges the Yangyang Development Fund. K.W. and T.T. acknowledge support from the JSPS KAKENHI (Nos. 21H05233 and 23H02052) and World Premier International Research Center Initiative (WPI), MEXT, Japan. C.L. acknowledges China Postdoctoral Science Foundation (No. GZB20230422)

## Author contributions

S.W., Z.S., C.L., and J.L. designed and supervised the experiment. Z. L., S.J. and Y.G. fabricated the devices. Y.L. and C.L. performed the STM/STS measurements. K.L., Q.S., L.L., X.L., D.G., Y.L., H.Z., C.L., J.J., T.L., G.C. analyzed the data. C.L., M.L. and J.L. performed theoretical studies. K.W. and T.T. grew the bulk hBN crystals. S.W. wrote the manuscript. All authors discussed the results and commented on the manuscript.

## Competing interests

The authors declare no competing financial interests.

**Data and Code Availability**

All data and code are available from the corresponding authors upon reasonable request.

**Materials & Correspondence**

Correspondence and requests for materials should be addressed to Jianpeng Liu, Can Li, Zhiwen Shi, and Shiyong Wang.

**Extended Data Figures**

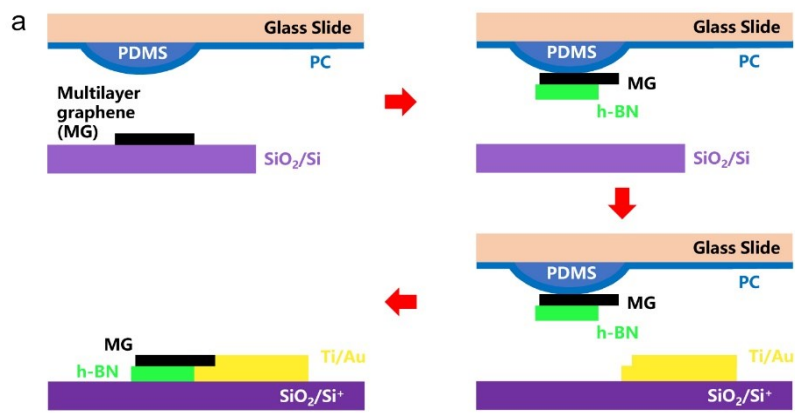

**Extended Data Fig. 1. Device fabrication procedure. a,** Schematic of the device fabrication processes. All the heterostructures were fabricated via the standard dry transfer method using a polycarbonate film placed on a polydimethylsiloxane stamp. We sequentially picked up layers from top to bottom and released them onto the substrate.

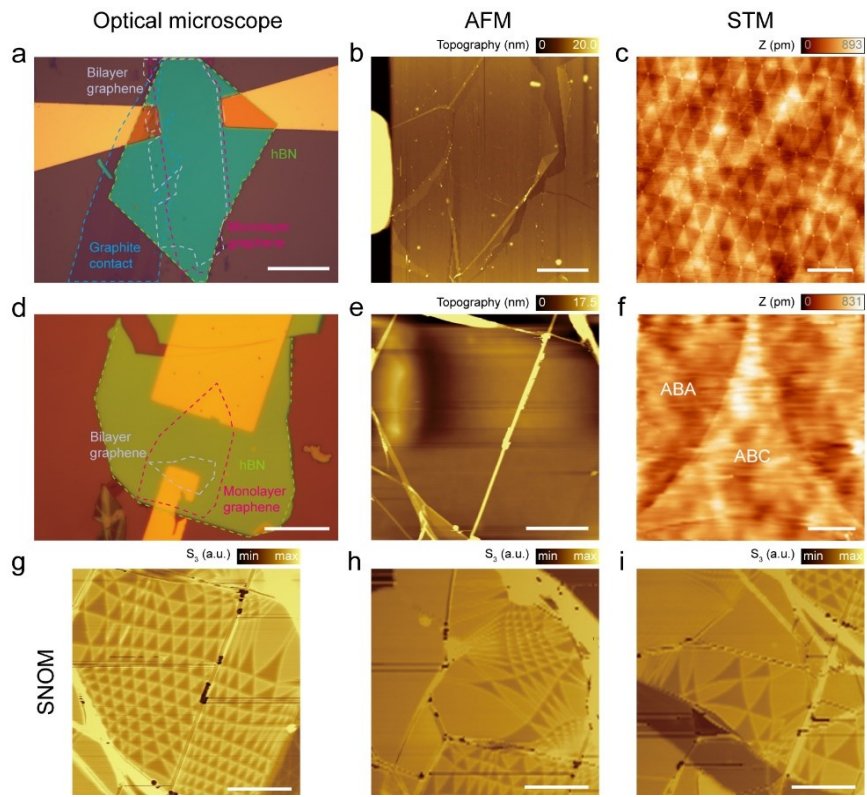

**Extended Data Fig. 2. Two Trilayer RG devices. a-c**, Optical micrograph, AFM, and STM images of trilayer RG device **1**. **d-f,** Optical micrograph, AFM, and STM images of trilayer RG device **2**. **e-f,** SNOM images of trilayer RG device **2**. A monolayer and a bilayer were sequentially picked up with a small twisting angle of 0.2°, resulting in alternating triangular domains of ABA and ABC stacking. Scale bars are 25um (a,d), 5 um (a, d,), 200nm (c), 1um (e, g, h, i) and 20 nm (f).

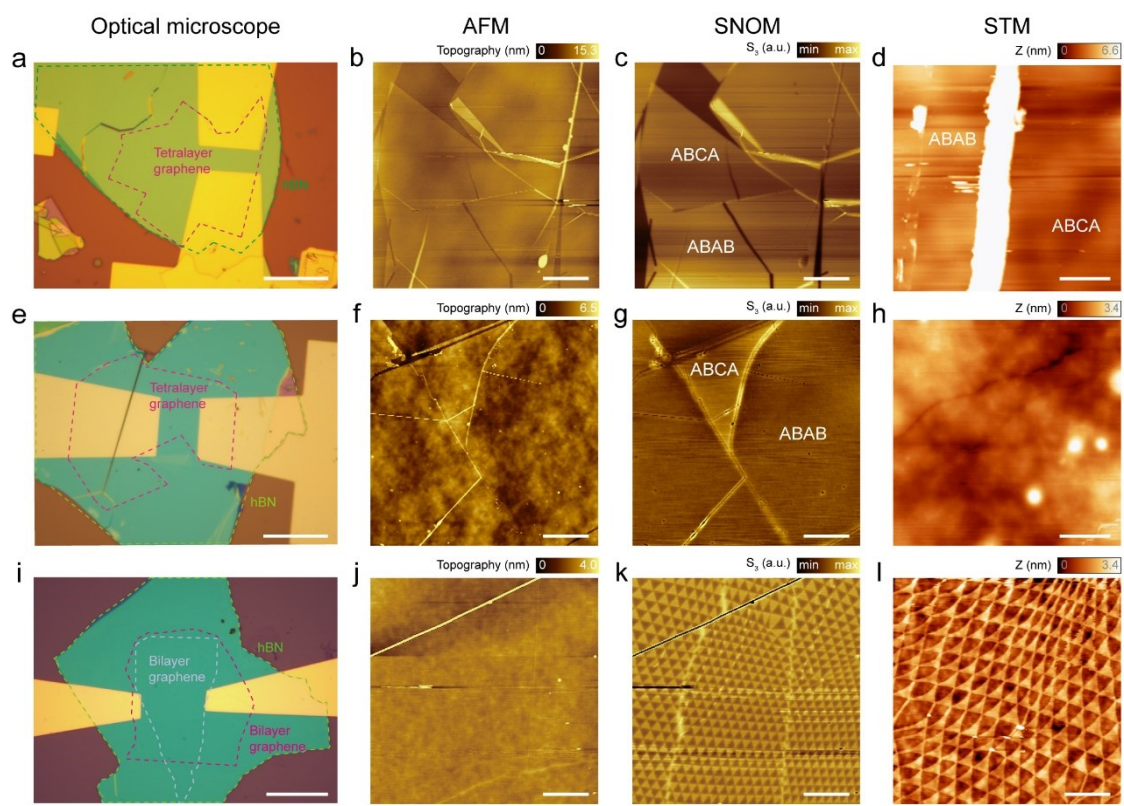

**Extended Data Fig. 3. Three tetralayer RG devices. a-d**, Optical micrograph, AFM, SNOM, and STM images of tetralayer RG device **3**. **e-h,** Optical micrograph, AFM, SNOM, and STM images of tetralayer RG device **4**. **i-l,** Optical micrograph, AFM, SNOM, and STM images of tetralayer RG device **5**. Device **3-4**: ABCA RG was obtained by screening naturally formed rhombohedral domains. Device **5:** Two bilayer flakes were sequentially picked up with a small twisting angle of 0.2°, resulting in alternating triangular domains of ABAB and ABCA stacking. Scale bars are 25um (a, e, i), 1 um (b, c, f, g, j, k), 100nm (d), 600nm (h) and 200 nm (l).

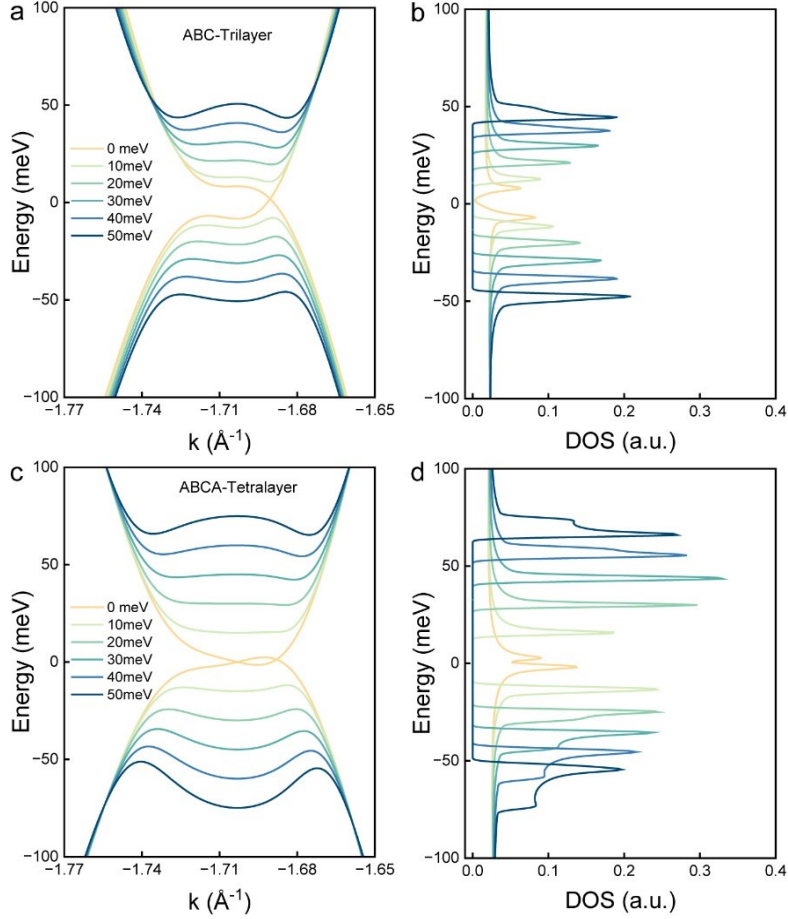

**Extended Data Fig. 4. Electronic structure of RG under electric field. a,b,** Tight-binding calculated band structure and density of states of trilayer RG under varying interlayer potential differences. **c,d,** Corresponding tight-binding calculations for tetralayer RG. The application of a perpendicular electric field induces a tunable bandgap in the electronic structure of both trilayer and tetralayer RG. This bandgap formation is accompanied by a significant flattening of the band edges, which is evident in the calculated band structures and DOS spectra.

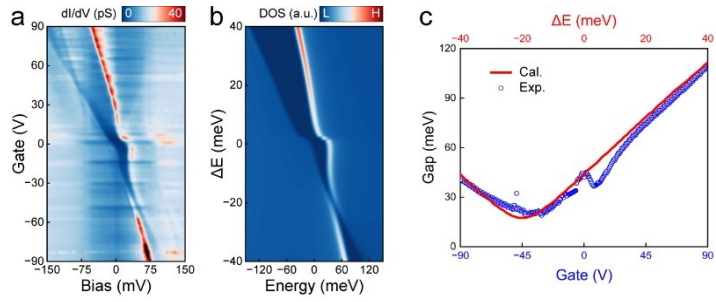

**Extended Data Fig. 5. Electronic structure of trilayer RG under tip electric field. a**, *dI/dV* spectra of trilayer RG as a function of back-gate voltage, showing an asymmetric gap opening caused by the tip-induced electric field. **b,** Tight-binding calculated DOS spectra under varying interlayer potential differences. **c,** Extracted gaps as a function of back-gate voltage. The gap is defined as the energy difference between CB bottom and VB top. An enhanced gap is observed at the charge neutrality point.

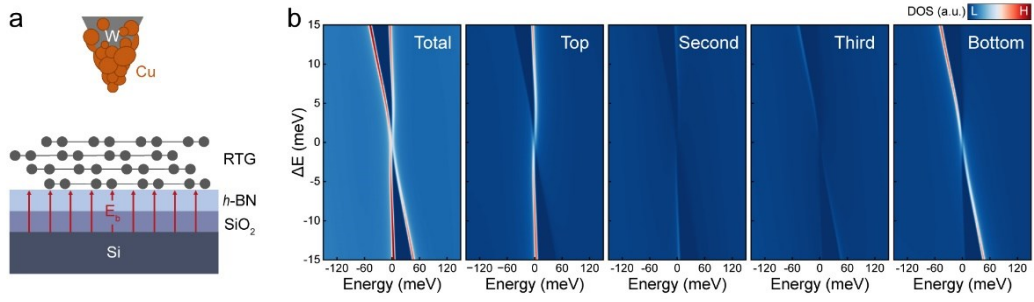

**Extended Data Fig. 6. Electric field-dependent electronic structure of tetralayer RG**. **a,** Schematic of the tetralayer RG setup using a copper STM tip. A silicon back-gate is employed to apply a perpendicular electric field. **b,** Tight-binding calculated DOS spectra under different interlayer potential differences. From left to right: total DOS, top-layer DOS, second-layer DOS, third-layer DOS, and bottom-layer DOS, respectively.

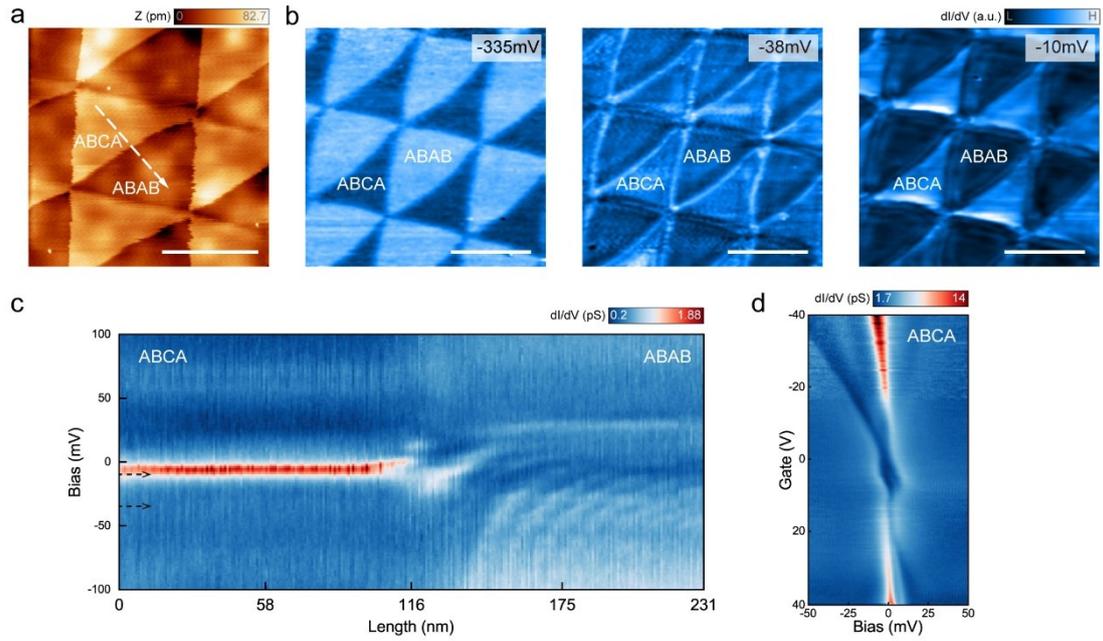

**Extended Data Fig. 7. Electronic structure of tetralayer RG device 5**. **a**, STM imaging ($V_{bias}$ = -400mV and $I$ = 10pA) showing alternative ABAB and ABCA domains. **b**, Differential conductance mappings taken at different biases as marked in (**c**). A one-dimensional edge states have been observed at the domain boundaries. **c**, Line d$I$/d$V$ spectra acquired along the dashed line in (a). **d**, d$I$/d$V$ spectra as a function of back-gate voltage, revealing a correlated gap at the charge neutrality point. Voltage modulations are 10mV (b), 2mV (c), 1mV(d). Scale bars are 200nm.

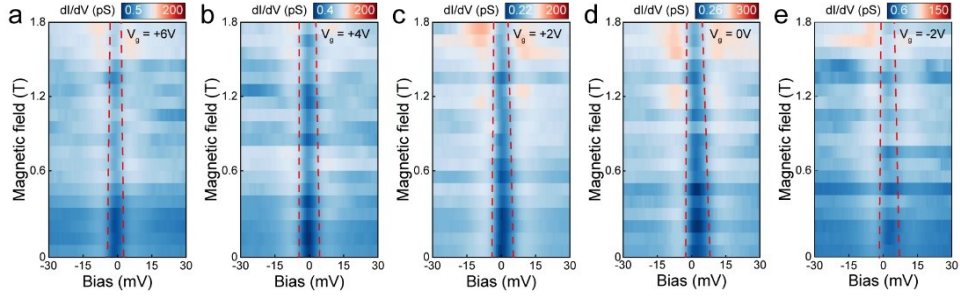

**Extended Data Fig. 8. Magnetic and doping induced insulating gap suppression in ABCA RG. a-e,** $dI/dV$ spectra taken at different back-gate voltages as an out-of-plane magnetic field. From left to right, the back-gate voltage is 6V, 4V, 2V, 0V and -2V, respectively. The dashed line indicates the gap edge. The gap is suppressed by a perpendicular magnetic field and charge carrier doping. Voltage modulations are 0.8mV.

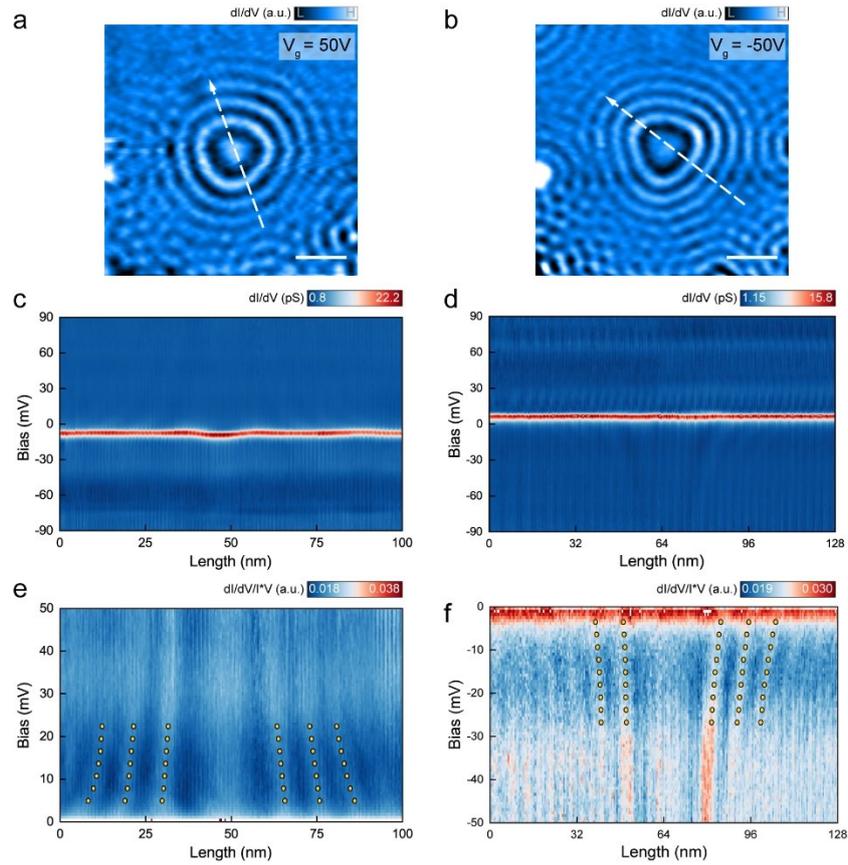

**Extended Data Fig. 9. Friedel Oscillations in tetralayer RG. a-b**, Large-scale d$I$/d$V$ mappings taken at back-gate voltage of 50V and -50V, respectively. Parameters: $V_{bias}$ = 10mV and $I$ = 300pA. **c-d**, Line dI/dV spectra measured along the corresponding white dashed lines in (a) and (b). **e-f**, Normalized line spectra d$I$/d$V$/(I/V) corresponding to (c) and (d), respectively. The dots mark the locations of the standing wave maxima. Voltage modulations are 5mV(a, b), 1mV(c, d, e, f). Scale bars are 30nm.